\documentclass[aps,prd,twocolumn,longbibliography,nofootinbib]{revtex4-1}
\usepackage{amsmath,amsfonts}

\begin{document}

\newcommand\be{\begin{equation}}
\newcommand\ee{\end{equation}}

\title{Gauge is more than mathematical redundancy}

\author{Carlo Rovelli}
\affiliation{Aix Marseille Universit\'e, CNRS, CPT, UMR 7332, 13288 Marseille, France.\\ Universit\'e de Toulon, CNRS, CPT, UMR 7332, 83957 La Garde, France.}
\date{\today}

\begin{abstract}\noindent
Physical systems may couple to other system through variables that are \emph{not} gauge invariant. When we split a gauge system into two subsystems,  the gauge-invariant variables of the two subsystems have \emph{less} information than the gauge-invariant variables of the original system; the missing information regards degrees of freedom that  express relations between the sub-systems.  All this shows that gauge invariance is a formalization of the relational nature of physical degrees of freedom.  The recent developments on boundary variables and boundary charges are clarified by this  observation. 
\end{abstract}

\maketitle

Gauge invariance is often described as convenient mathematical redundancy.  This description is misleading, because it hides the reason for which the world appears to be well described by gauge theories, such as Yang-Mills theory and general relativity \cite{Rovelli2014a,Teh2013,Amaral2014,Gomes2019,Gomes2019a,Freidel2019,Vanrietvelde2018}.

The best way to appreciate the physical meaning of gauge is to consider splitting a system into components.  There are various ways in which this could be done.  

As a first example, consider a field theory defined by various interacting fields. We can separate one field and consider it in isolation, neglecting its interaction with the other fields.   Consider for instance QED. The corresponding classical field theory is defined by the action
\be
S[A,\psi]=\int {\frac14} F[A]^2+\bar\psi\,\slash\!\!\!\!D[A]\psi \label{action}
\ee 
written in terms of the Dirac field $\psi$ and the Maxwell potential $A$. Here $F[A]$ and $D[A]$ are curvature  and covariant derivative of $A$.  If we neglect the Dirac field, the electromagnetic field alone is described by the first term  of \eqref{action} alone, the theory is invariant under the gauge 
\be
A\to A+d\lambda  \label{gauge}
\ee
and (if spacetime is topologically trivial) the electric and magnetic fields capture all the gauge invariant degrees of freedom. But the way electromagnetism couples to the Dirac field in \eqref{action} is via the local interaction term of the Lagrangian density
\be
   I= \bar\psi\,\slash\!\!\!\!A\psi 
\ee
which depends on the \emph{non} gauge-invariant variable $A$.  Therefore $A$ in this formalism is not  mathematical redundancy: it is the variable of electromagnetism that can couple locally to the Dirac field. 

As a second example, consider a non abelian Yang-Mills theory, in a truncation where we split spacetime into discrete cells.  We can represent every cell $c$ of spacetime as a node of a lattice and the theory can be formulated using Wilson's formalism of lattice gauge theory.  The theory can be discretised in terms of group variables $U_{cc'}$ associated to every couple of adjacent cells $c$ and $c'$.  The discrete version of the Yang-Mills gauge is the transformation 
\be
U_{cc'}\to \Lambda_cU_{cc'}\Lambda_{c}^{-1}    \label{gauge2}
\ee
where $\Lambda_c$ are arbitrary group elements.   The gauge invariant variables are the Wilson loops
\be
U_{cc'}U_{c'c''}U_{c''c'''}....U_{c^nc}.
\ee
Now let us split a large spacetime region $\Sigma$ formed by many cells into two subregions  $\Sigma_1$ and  $\Sigma_2$.  The variables $U_{cc'}$ split into \emph{three} groups: those where $c$ and $c'$ belong (i) both to the first subregion, (ii) both to the second subregion, and (iii) to  distinct subregions. Let us call these variables respectively $U^{1}, U^{2}$ and $U^{12}$.  The first and the second region are described by the variables $U^{1}$ and $U^{2}$ respectively, whose gauge invariant functions are the loop variables entirely in the first or in the second region. These variables are \emph{not} sufficient to describe the degrees of freedom of the full region $\Sigma$, because the $U^{12}$ variables are missing. To couple the two regions we need these extra variables;  notice that the variables $U^{12}$ are not gauge invariant. They express the change of internal frame from one region to the other.  They are the handles through which the two subsystems couple. Once again, non-gauge-invariant variables are ways through which a system can couple to another system, a spacetime region to another spacetime region. (On this, see also \cite{Gervais1976,Gervais1976,Balachandran2013,Donnelly2016,Geiller2018,Speranza2018,Freidel2017,Freidel2018,Freidel2019}.)

As a third example, consider a non rotating black hole in pure general relativity.\footnote{I thank Laurent Freidel for pointing out this example.}   As well known, there is a single solution up to gauges that describes such a black hole in the theory.  This is given, in one coordinate system, by the Schwarzschild metric.  We do not associate a position or a velocity (as we do for a particle in Minkowski space) to a black hole solution in general relativity, because a Schwarzschild black hole whose coordinate position is changing in time in a given coordinate system is a solution of the Einstein's equation that is gauge equivalent to a static one.   Therefore the position and velocity of a black hole are pure gauge in this sense. 

But does this mean that astrophysical black holes have no position or velocity?  Of course not: astrophysical black holes have positions and velocities.  How come so, if there is only a single solution of Einstein equations describing a non rotating stationary black hole?  The answer is of course that in the universe there are other physical components than those entering  he Schwarzschild solution alone, and in the coupled system formed by all the various components of the universe the position and the velocity of a black hole can be appropriately defined with respect to other objects: say for instance with respect to its  its host galaxy.  There is nothing particularly deep in all that, but notice something: the position and the velocity of the black hole that were \emph{gauge variables} when considering the hole alone, become physically meaningful variables when the black hole is coupled with something else.  Again, we see that gauge variables are handles through which a system  couples with something else. 

In all these cases, we see what is behind gauge invariance: the fact that physical degrees of freedom are often not attached to specific entities or locations, but bridge between these. Gauge-invariant quantities can be definied by coupling gauge non-invariant quantities from different systems. 

Notice that this implies that  in some sense one can ÒmeasureÓ a non gauge-invariant quantity of a system as long as it is relative to another system. When the measuring apparatus couples to a physical system to mesaure something, the coupling may be gauge invariant under a common gauge transformation on the system and the apparatus, but \emph{the measured quantity pertaining to the system can be a non gauge-invariant variable of the system alone}.   For nice examples of applications of this perspective, see \cite{Leader2014,Alkofer2014}.

In recent years there has been a flourishing of interest on boundary charges, and boundary degrees of freedom and in particular asymptotic charges and asymptotic degrees of freedom \cite{Strominger2014,Kapec2017,Kapec2017a,Campiglia2017,Nande2017}. The discussion above clarifies the physics underlying this phenomenology.   To see this, consider again the case of lattice gauge theory on a region $\Sigma$ split into two subregions $\Sigma_1$ and $\Sigma_2$.  Suppose we study the two regions $\Sigma_1$ and $\Sigma_2$ \emph{separately}.  If we consider only the gauge invariant variables of them, and we neglect the  $U^{12}$ variables, we are clearly missing something relevant for physics.  Hence we must consider these variables as well.  But they sit on the boundary and are non-gauge invariant variables when considering one of the two regions alone, with its boundary.   Hence, if we do not want to miss relevant physics, we must not neglect boundary variables even if they are not gauge invariant.   

What do they represent?  They capture aspects of the way the region can interact with whatever is on its boundary.  In particular, any measure on its boundary is an interaction with the region.  The measuring apparatus may couple with the region at the boundary, the coupling may be gauge invariant under a common gauge transformation on the system and the apparatus, but \emph{the measured quantity pertaining to the bounded region can be a non gauge-invariant variable of the field in that region}.   

The same is true for asymptotic symmetries and asymptotic observables: non gauge-invariant asymptotic variables are physically measurable because in a physical measurement an observer can interact with these variables. One again: the interaction is gauge invariant, but the observed variable is not.  In the above example of the black hole, the black hole can be given a position and a velocity by shifting or boosting the solution at infinity. 

In summary, considering gauge invariance as mathematical redundancy obscures its physical significance: many physical quantities express relations between distinct systems.

\end{document}